\documentclass[a4paper]{jpconf}
\usepackage{graphicx}
\usepackage{amsmath}

\def\om{\omega}

\renewcommand{\lsim}{\stackrel{\scriptstyle <}{\phantom{}_{\sim}}}
\newcommand{\gsim}{\stackrel{\scriptstyle >}{\phantom{}_{\sim}}}

\begin{document}

\title{Hyperon puzzle and the RMF model with scaled hadron masses and coupling constants}

\author{E~E~Kolomeitsev$^1$, K~A~Maslov$^2$, D~N~Voskresensky$^2$}

\address{
$^1$Matej Bel University, SK-97401 Banska Bystrica, Slovakia\\
$^2$National Research Nuclear  University (MEPhI), 115409 Moscow, Russia}

\ead{Evgeni.Kolomeitsev@umb.sk}

\begin{abstract}
The equation of state of cold baryonic matter is studied within a
relativistic mean-field model with hadron masses and coupling
constants depending on a scalar field.
We demonstrate that if the effective nucleon mass stops to decrease with a density increase at densities $n>n_*>n_0$, where $n_0$ is the nuclear saturation density, the equation of state stiffens for these densities and the limiting neutron star mass increases.
The stabilization of the nucleon mass can be realised if in the equation of motion for the scalar mean-field there appear a term sharply varying in a narrow vicinity of the field value corresponding to the density $n_*$. We show several possible realizations of this mechanism getting sufficiently stiff equations of state.
The appearance of hyperons in dense neutron star interiors is
accounted for. The obtained equations of state remain sufficiently
stiff if the reduction of the $\phi$ meson mass is incorporated. Thereby, the hyperon puzzle can be resolved.
\end{abstract}

\section{Introduction}

In this contribution we present the results of our studies of the equation of state (EoS) for the cold nuclear matter. We work on the extension of a relativistic mean-field (RMF) model~\cite{KVOR} including a possibility for scaling of hadron masses and meson-baryon coupling constants. As in~\cite{KVOR}, all hadron masses are  scaled in the same way, whereas the coupling constants are scaled differently.

The current challenge is to construct such an EoS that can describe a neutron star (NS) with the mass of $(2.01\pm 0.04) M_\odot$, where $M_\odot$ is the mass of the Sun, which was recently observed, cf. ~\cite{Demorest:2010bx,Antoniadis:2013pzd}. For that the EoS should be rather stiff.
Simultaneously the EoS has to respect the constraint derived from the flow of particles produced in heavy-ion collisions~\cite{HICflow}. That oppositely requires the EoS to be rather soft. The additional complication in the description of heavy NSs arises if hyperons are included into consideration, since their presence softens the EoS, cf.~\cite{Hypprob}.

\section{The RMF model with scaling}\label{sec:model}
The nuclear energy-density functional of the RMF model including  members of the baryon SU(3) octet ($B=N,\Lambda,\,\Sigma,\,\Xi$) which interactions are mediated by the scalar meson ($\sigma$) and non-strange members of the vector-meson SU(3) nonet $V=\om,\rho,\phi$, is
\begin{eqnarray}
E[f] =  \sum_B \int_0^{p_{{\rm F},B}} \frac{p^2 dp}{\pi^2} \sqrt{p^2 + m_B^{*2}(f)}
+\frac{m_N^4 f^2}{2 C_\sigma^2 } + U(f)  +\sum_V \frac{C_V^2\widetilde{n}_{V}^2}{2m_N^2\eta_V(f)}.
\label{Efunc}
\end{eqnarray}
Here $p_{{\rm F},B}=(3\pi^2 n_B)^{1/3}$ is the baryon Fermi momentum with $n_B$ standing for the  density of baryon $B$. The effective baryon density $\tilde{n}_V$, being a source for the mean-field of meson $V$, is defined as $\widetilde{n}_{V}=\sum_B x_{V B}q_V n_B$, where  $q_V$ stands for the baryon charge for $V=\om$, the baryon third isospin projection for $V=\rho$, and the baryon strangeness for  $V=\phi$. The effective baryon mass $m_B^*$ differs from the vacuum mass $m_B$ because of the coupling to the scalar field, $m_B^*(f)=m_B-x_{\sigma B}m_N\,f$, where $f=g_{\sigma N} \sigma/m_N$ is the dimensionless  scalar field, and $g_{\sigma N}$ is the $\sigma N$ coupling constant.
The functions $\eta_V(f)$ are the key ingredients, which differ our approach from other RMF models. These functions stem from the ratios of scalings of vector-meson masses to scalings of hadron-nucleon coupling constants and are tuned in our approach to render the best agreement of the resulting EoS with empirical constraints.  The scalar field is found from the minimization of the functional (\ref{Efunc}) with respect to $f$: $\delta E[f]/\delta f=0$.
The coupling constants $C_{\sigma,\om\,\rho}$ are to be determined from the saturation properties of the nuclear matter, therefore we have $x_{\sigma N}=x_{\om N}=x_{\rho N}=1$. Since the $\phi$ meson does not couple to a nucleon, if  the strict Okubo-Zweig-Iizuka selection rule is applied, i.e.  $x_{\phi N}=0$, the coupling constant $C_\phi$ is related to the coupling of the $\om$ meson as $C_\phi=C_\om m_\om/m_\phi$, where  $m_\om = 783\,{\rm MeV}$, $m_\phi = 1020\,{\rm MeV}$. The coupling constants of vector mesons to hyperons are chosen traditionally according to the quark SU(6) symmetry
\begin{eqnarray}
x_{\om \Lambda} =x_{\om \Sigma}= 2x_{\om \Xi} = {\textstyle\frac{2}{3} } \,, \,\,\,
x_{\rho \Sigma} = 2 x_{\rho \Xi} = 2\,,\,\,\,
x_{\phi \Lambda} = x_{\phi \Sigma} = x_{\phi \Xi} = - {\textstyle\frac{\sqrt{2}}{3} }\,,\,\,\,
x_{\rho\Lambda}=x_{\phi N} = 0.
\label{gHm}
\end{eqnarray}
The coupling constants of hyperons with the scalar mean field are
derived from the hyperon binding energies $\mathcal{E}_{\rm bind}^{H}$ in isospin-symmetric matter (ISM) at the saturation density $n=n_0$ given by~\cite{KVOR}
$
\mathcal{E}_{\rm bind}^{H}(n_0) = C_{\omega}^2 m_N^{-2}x_{\omega H}n_0 - m_N+m_N^{*} (n_0)\,,
$ and the empirical values
$ \mathcal{E}_{\rm bind}^{\Lambda }(n_0) = -28$\,MeV,
$\mathcal{E}_{\rm bind}^{\Sigma }(n_0) = 30$\,MeV and
$\mathcal{E}_{\rm bind}^{\Xi }(n_0) = -15$\,MeV.
The repulsive $\Sigma$ potential prevents the appearance of $\Sigma$ hyperons in all models considered below.

For the description of the NS one should add to (\ref{Efunc}) the energies of leptons.
The NS composition is governed by the weak processes bringing the matter in the $\beta$-equilibrium and relating chemical potential of all species to the electron and neutron chemical potentials. Together with the electroneutrality condition these relations allow to express the densities of all baryonic and leptonic species as functions of the total baryon density $n=\sum_B n_B$. The pressure of the NS matter follows then as $P(n)=n\partial E/\partial n-E$.
With $P = P(E)$ at hand, the NS masses and radii are calculated with the help of the Tolman-Oppenheimer-Volkoff (TOV) equation. In the crust region we match smoothly our EoS with the BPS EoS~\cite{BPS}.

\section{A novel mechanism how one can stiffen the EoS in an RMF model}

With appropriate choices of the scaling functions $\eta_V(f)$, and $U(f)$ and, perhaps, also with a modification of $m_B^*(f)$, our model can be reduced to various traditional RMF models. For example, putting \begin{eqnarray}
\eta_\om(f) = \eta_\rho(f) = 1, \quad U(f) = m_N^4 (b f^3/3 + c f^4/4)
\label{NLWscale}
\end{eqnarray}
we recover the extension of the original Walecka model proposed by Boguta and Bodmer~\cite{BB}, the so-called non-linear Walecka (NLW) model, which allows to fit satisfactory the nuclear saturation properties, which we chose here for an illustration as follows:
the saturation density $n_0=0.16\,{\rm fm}^{-3}$, the binding energy per nucleon $\mathcal{E}_0=-16$\,MeV, the compressibility modulus $K=250$\,MeV, the effective nucleon mass $m_N^*(n_0)/m_N=1-f_0=0.8$, and the symmetry energy $E_{\rm sym}=30\,$MeV.
For this choice of parameters and scalings (\ref{NLWscale})  the limiting mass of a NS is $1.92~M_\odot$ that is below the experimental limit.

In~\cite{MKV-cut} we proposed a simple method how one can stiffen the EoS in an RMF model.
The strategy is to reduce the growth of the scalar field above a  value $f_{\rm cut}$. As argued in~\cite{MKV-cut}, this can be achieved by adding to the original energy-density functional (\ref{Efunc}) a function of the scalar field, which is vanishingly small for $f \sim f(n_0) < f_{\rm cut}$ and is sharply increased  for $f\sim f_{\rm cut}$, so that its second derivative becomes sufficiently large. Then, the self-consistent solution of the  equation for $f$ does not let the function $f(n)$ to grow significantly above $f_{\rm cut}$ within a broad interval of densities for $n>n_{*}$, where $n_*$ is determined by $f_{\rm cut}$.
We will illustrate this mechanism at  hand of the following example: we replace $U$ in (\ref{NLWscale}) by $U(f)+\Delta U(f)$, where
\begin{eqnarray}
\Delta{U}(f)=\alpha\ln[1+\exp(\beta(f-f_{\rm cut}))]\,,\quad \alpha= m_\pi^4\,,\quad \beta=120.
\label{Umod}
\end{eqnarray}
We will refer to such an extension as the ``$\sigma$-cut scheme'' and call the model as the NLWcut model.
Applying the method to models with different values of $f_0$ it is convenient to parameterise
$f_{\rm cut}=f_0+c_\sigma(1-f_0)$, and allow different values of $c_\sigma$ to vary. The NLW model is recovered from the NLWcut model for $c_\sigma\gsim 1$ corresponding to $f_{\rm cut}\gsim 1$.

\begin{figure}
\centering\includegraphics[width=0.98\textwidth]{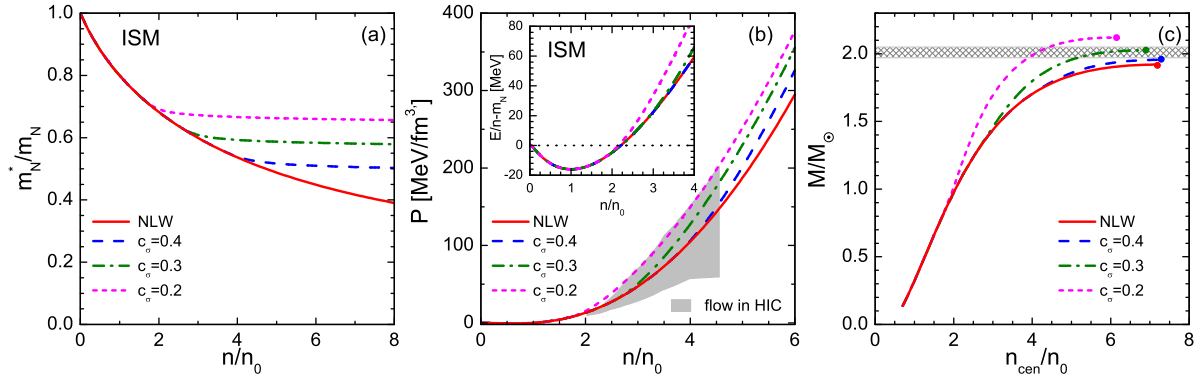}
\caption{Results for the NLW and NLWcut models.
Panel (a): the nucleon effective mass in the ISM as a function of the nucleon density.
Panel (b):  Pressure $P$ as a function of the nucleon density in the ISM for various values $c_{\sigma}$; the insertion shows the binding energy per nucleon. The shadowed area is the constraint on the pressure from the particle flow in heavy-ion collisions (HIC) obtained in~\cite{HICflow}.
Panel (c): The NS mass as a function of the central density. The hatched band denotes the uncertainty in the value of the measured heaviest NS mass~\cite{Antoniadis:2013pzd}.
}
\label{fig:sig-cut}
\end{figure}

On panel (a) in Fig.~\ref{fig:sig-cut} we show the effective nucleon mass $m_N^*$ as a function of the nucleon density in ISM. We see that the function $m_N^*(n)$ flattens off for densities $n>n_{*}\simeq 1.9\,n_0$, $3.0\,n_0$, and $4.0\,n_0$ for $c_\sigma=0.2$, $0.3$, and $0.4$, respectively.
The influence of the quenching of the nucleon-mass decrease on the EoS is illustrated on panel (b) in Fig.~\ref{fig:sig-cut}, where we show the total pressure $P(n)$ and the binding energy per nucleon as functions of $n$ in the ISM. We observe that the replacement (\ref{Umod}) leads to a sizable increase of the pressure for $n>n_{*}$. The shadow region shows the constraint on the pressure extracted from the particle flow in HIC~\cite{HICflow}. We see that the EoSs in the NLWcut models satisfy this constraint for $c_\sigma\gsim 0.2$.
In difference from the pressure, the energy density and corresponding binding energy increase much weaker after the inclusion of the new term~(\ref{Umod}). The NS masses as functions of the central density  are shown on panel (c) in Fig.~\ref{fig:sig-cut}.
The stiffening of the EoS reflects in an increase of the NS mass at a given central density. The proposed $\sigma$-cut scheme allows us to shift the maximum NS mass from $1.92\,M_\odot$ for the original NLW model to $1.96\,M_\odot$, $2.03\,M_\odot$, and $2.12\,M_\odot$ for the NLWcut models with $c_\sigma=0.4$, $0.3$, and $0.2$, respectively.

\section{KVOR, KVORcut and MKVOR models}

In~\cite{KVOR} the model KVOR was proposed with scaling functions $\eta_\rho > 1$ and $\eta_\omega<1$ for $n>n_0$, which were chosen to shift the threshold of direct Urca (DU) process, $n\to p+e+\bar{\nu}$, to higher densities (via the $\eta_\rho$ tuning) and to increase the maximum neutron star mass (via the $\eta_\om$ tuning). The KVOR model exploits the  saturation parameters $m_N^*(n_0)/m_N=0.805$, $K=275$\,MeV, $J_0=32$\,MeV, and other parameters being the same as for the NLW model discussed above.  The KVOR EoS was shown in~\cite{Klahn:2006ir} to satisfy appropriately the majority of experimental constraints known by that time.  Hyperons were not
included in the original KVOR model. However even without hyperons the KVOR EoS yields the maximum mass $2.01 M_{\odot}$ that fits the
constraint~\cite{Antoniadis:2013pzd} only marginally.

\begin{figure}\centering
\includegraphics[width=0.98\textwidth]{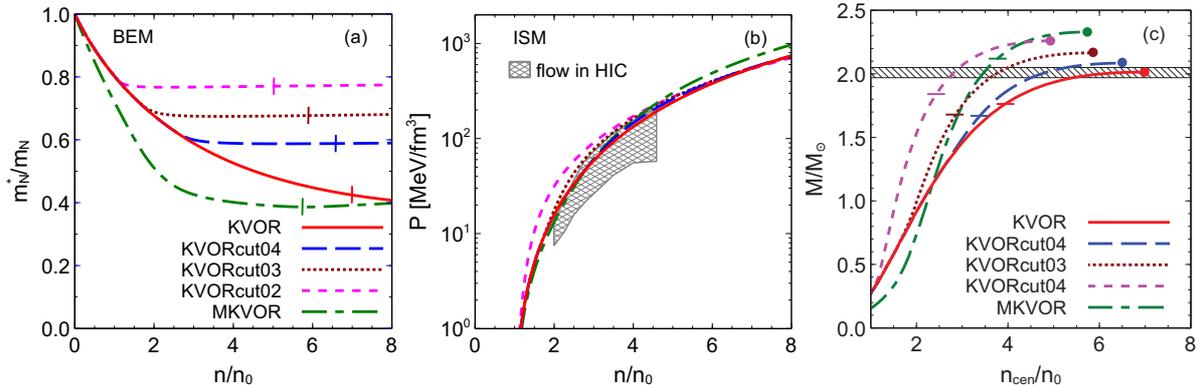}
\caption{Panel (a): The nucleon mass as a function of the density in KVOR, KVORcut and MKVOR models in BEM. Vertical bars indicate central densities in the stars with the maximum
masses for the corresponding model.
Panel (b): Pressure as a function of the nucleon density for ISM. Double-hatched area is the constraint from the particle flow in heavy-ion collisions~\cite{HICflow}.
Panel (c): The NS mass versus the central density. Bold dots show maximum available masses. The band shows uncertainty range of the measured mass of PSR J0348+0432~\cite{Antoniadis:2013pzd}. Horizontal dashes show the DU thresholds.}
\label{fig:KVORcut}
\end{figure}

We discuss now several extensions of the KVOR model~\cite{KVOR}, which can better fulfill the modern constraints on the nuclear EoS. To stiffen the EoS we use the method described in the previous section but now will change not the potential $U(f)$ but the $\eta_\om(f)$ function as
\begin{eqnarray}
\eta_\om^{\rm KVOR}(f)\to \eta_\om^{\rm KVOR}(f)+\frac{a_\omega}{2} \big[1 + \tanh(b_\om (f - f_{{\rm cut},\om}))\big]\,.
\end{eqnarray}
We will call such extensions the KVORcut models. We consider three choices of parameters $a_\om$, $b_\om$ and $f_{{\rm cut},\om}$:
KVORcut04 with $a_\om=-0.5$, $b_\om=55.76$, $f_{{\rm cut},\om}=0.454$\,;
KVORcut03 with $a_\om=-0.5$, $b_\om=46.78$, $f_{{\rm cut},\om}=0.365$\,;
KVORcut02 with $a_\om=-0.2$, $b_\om=74.55$, $f_{{\rm cut},\om}=0.249$\,.
In~\cite{MKV-short} we introduced another RMF model with scaling --- the MKVOR model --- where the mechanism of the $f$ stabilization is implemented not in the $\eta_\om(f)$ function but in the $\eta_\rho(f)$ function, which rises for $f<0.5$ similarly to the KVOR model but drops fast for $f > 0.52$. Therefore in the MKVOR model the nucleon mass decreases with the density increase in the ISM matter stronger than in the $\beta$-equilibrium matter (BEM). All other scaling functions we chosen anew to satisfy better various constraints on the EoS form microscopic calculations, see~\cite{MKV-short} for details.

The nucleon effective masses for  KVORcut and MKVOR models are shown in Fig.~\ref{fig:KVORcut}, panel (a), as functions of the baryon density for BEM. In all models the
effective hadron masses first decrease with the density and then saturate at
some approximately constant values. The pressure as a function of the nucleon density in the ISM is shown on panel (b) in Fig.~\ref{fig:KVORcut}. We see that pressure for the KVORcut04 model
is only slightly higher than that of the KVOR one. The curve for the KVORcut03 model goes on top of the upper boundary of the experimentally allowed region.
The MKVOR-model curve goes through the region for $n < 4n_0$ but escapes it at higher density. Oppositely, the curve for the KVORcut02 model lies above the region for $n < 3.5n_0$ but enters it for higher $n$.
Panel (c) in Fig.~\ref{fig:KVORcut} demonstrates the NS mass as a function of the
central density for our EoSs. The band shows the uncertainty range of the masses for PSR~J0348+0432~\cite{Antoniadis:2013pzd} which is fulfilled by all models. We see that for the KVORcut models the lesser the value $f_{\om}$ is chosen, the larger the value of the maximum mass is and the smaller the central density corresponding to $M_{\rm max}$ becomes.
The MKVOR EoS is the stiffest among  the EoSs considered here. Dashes on the right panel indicate the DU thresholds. The MKVOR model yields the highest DU threshold density.
Thus, the KVORcut03 and MKVOR models are the most promising models for the simultaneous fulfillment of the particle-flow and NS-mass constraints.

\section{Inclusion of hyperons}

We discuss now the results of the hyperonization phase transition on the EoS within our RMF models.
We will consider three schemes for the inclusion of hyperons:\\
(i) The hyperons are included with the vector-meson coupling parameters $x_{\om H}$ and  $x_{\rho H}$ according to (\ref{gHm}) and the scalar coupling parameters $x_{\sigma H}$ constrained by the hyperon binding energies as outlined in section~\ref{sec:model}. We switch off the hyperon coupling to $\phi$ mesons, putting $x_{\phi H}=0$.
The models with hyperons obtained within such a scheme are labeled as the KVORH, KVORHcut03 and MKVORH models.\\
(ii) Next we incorporate the $\phi$-meson mean field and assume the very same scaling of its
mass as for all other hadrons, i.e., $m_\phi^*(f) = m_\phi(1-f)$, but with the unscaled hyperon-$\phi$ coupling constants, which gives the scaling function $\eta_\phi(f) = (1-f)^2$.
The resulting models are denoted as KVORH$\phi$, KVORHcut03$\phi$ and MKVORH$\phi$.\\
(iii) In addition to the $\phi$-meson mass scaling we take into account a possible change of the $x_{\sigma H}$ parameters with an increase of density. We assume these parameters being unchanged for densities $n\lsim n_0$ and decreasing for higher densities, so that they reach and stay zero for the baryon density $n > n_{cH}$, where $n_{cH}$ are critical densities for hyperonization.
With such a parametrization we will exploit vacuum masses of the hyperon $H$ for $n > n_{cH}$. Such extensions of the models are call  KVORHcut03$\phi\sigma$ and MKVORH$\phi\sigma$ models.

As known, the inclusion of hyperons results usually in a substantial decrease of
the maximum NS mass. The difference between NS masses
with and without hyperons proves to be so large for reasonable hyperon fractions
in the standard RMF approach that in order to solve the puzzle one
needs to start with a very stiff EoS without hyperons that would be in odds with microscopical calculations and with the constraints from heavy-ion collisions.

\begin{figure}\centering
\includegraphics[width=0.98\textwidth]{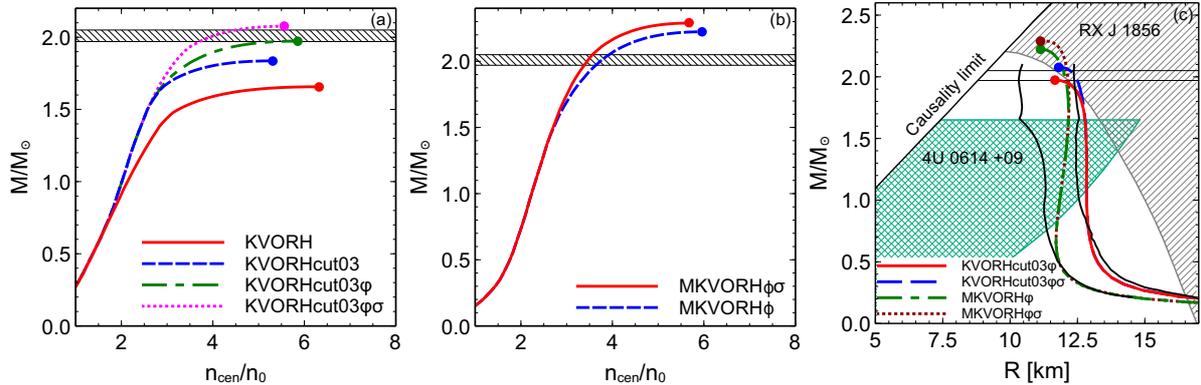}
\caption{Panels (a) and (b): NS mass versus the central baryon density for various models with hyperons. Panel (c): Mass-radius relation for KVORcut03 and MKVOR models including hyperons and constraints from thermal radiation of the isolated NS RX J1856~\cite{Trumper} and from quasi-periodic oscillations in the LMXBs 4U 0614+09~\cite{Straaten}. The black thin lines show the radius constraint from the analysis~\cite{Lattimer:2012nd}.
The horizontal band shows the uncertainty range of masses for PSR~J0348+0432~\cite{Antoniadis:2013pzd}. }
\label{fig:hyper}
\end{figure}

The NS masses as functions of the central baryon density for various models with hyperons are shown in Fig.~\ref{fig:hyper}.  Comparing with Fig.~\ref{fig:KVORcut} we see that the appearance of hyperons reduces the maximum NS mass for the  KVORH and KVORHcut03 models
by $0.35\,M_\odot$ and $0.33\,M_\odot$, respectively, bringing them below the empirically known pulsar mass. We find also that hyperons cannot be incorporated
in the MKVOR model, since the solution for $f(n)$ does not exist for all densities.
Including the $\phi$-meson with an $f$-dependent mass and scaling of $\sigma H$ coupling constants (schemes (ii) and (iii) above) we obtain maximum NS mass of
$1.97\,M_\odot$ for KVORHcut03$\phi$ that marginally agrees with the observational
constraint and $2.07\,M_\odot$ for KVORHcut03$\phi\sigma$ that fully fulfils the maximum NS mass constraint. For the extensions of the MKVOR models we find the maximum mass of
$2.22\,M_\odot$ for MKVORH$\phi$ and $2.29\,M_\odot$ for MKVORH$\phi\sigma$; both values are well above the observational maximum NS mass constraint.

Mass-radius relations for KVORHcut03$\phi$, KVORHcut03$\phi\sigma$, MKVORH$\phi$ and
MKVORH$\phi\sigma$ EoSs are shown on panel (c) in Fig.~\ref{fig:hyper}. We see that the MKVORH$\phi$ and MKVORH$\phi\sigma$ EoSs satisfy all presented constraints, whereas the KVORHcut03$\phi$ and KVORHcut03$\phi\sigma$ EoSs render star radii slightly larger than the corridor of allowed values (shown by black thin lines) extracted in~\cite{Lattimer:2012nd} from the analysis of empirical data.

\section{Conclusion}

We presented a novel mechanism of stiffening the EoS within an RMF model with hadron masses and coupling constants depending on the scalar field. The key idea is to quench the growth of the scalar field and correspondingly a decrease of the nucleon mass with a density increase.
Three different approaches are outlined where the quenching is achieved by the modification of the $\sigma$ mean-field self-interaction or by changing in the vector ($\omega$) and isovector ($\rho$) sectors of the energy density functional. Within these schemes we discuss the $\sigma$-cut versions of the traditional non-linear Walecka model, the $\omega$-cut versions of the KVOR model proposed in~\cite{KVOR} (KVORcut models) and a new model MKVOR combining various schemes.
The hyperons are included and it is demonstrated that KVORcut and MKVOR models can support sufficiently heavy neutron stars if a decrease of the $\phi$-meson mass or the hyperon-$\sigma$ coupling constants are incorporated.

This work was supported by the Ministry of Education and Science
of the Russian Federation (Basic part), by the Slovak Grants No.
APVV-0050-11 and VEGA-1/0469/15, and by ``NewCompStar'', COST
Action MP1304. E.E.K and K.A.M. acknowledge the hospitality of Joint Institute of Nuclear Research in Dubna (Russia), where part of this work was done. E.E.K acknowledges the support
by the Plenipotentiary of the Slovak Government to JINR Dubna.


\section*{References}
\begin{thebibliography}{99}
\bibitem{KVOR}
Kolomeitsev E E  and Voskresensky D N  2005
{\it Nucl.\ Phys.\ A} {\bf 759} 373.

\bibitem{Demorest:2010bx}
Demorest P \etal 2010
{\it Nature} {\bf 467} 1081.

\bibitem{Antoniadis:2013pzd}
Antoniadis J \etal 2013
{\it Science} {\bf 340} 6131.

\bibitem{HICflow}
Danielewicz P, Lacey R and Lynch W G 2002
{\it Science} {\bf 298} 1592.

\bibitem{Hypprob}
Schaffner-Bielich J 2010 {\it Nucl. Phys. A} {\bf 835} 279.

\bibitem{BPS}
Baym G, Pethick C and Sutherland P 1971
{\it Astrophys. J.} {\bf 170} 317.

\bibitem{BB}
Boguta J and Bodmer A R 1977 {\it Nucl. Phys. A} {\bf 292} 413.

\bibitem{MKV-cut}
Maslov K A, Kolomeitsev E E and Voskresensky D N 2015 arXiv:1508.03771

\bibitem{Klahn:2006ir}
Kl\"ahn T \etal 2006
{\it Phys.\ Rev.\ C} {\bf 74} 035802.

\bibitem{MKV-short}
Maslov K A, Kolomeitsev E E and Voskresensky D N 2015 {\it Phys. Lett. B} {\bf 748}  369

\bibitem{Trumper}
Tr\"umper J E \etal 2004
{\it Nucl.\ Phys.\ B (Proc.\ Suppl.)} {\bf 132} 560.

\bibitem{Straaten} van Straaten S et al. 2000
{\it Astrophys. J.} {\bf 540} 1049.

\bibitem{Lattimer:2012nd}
Lattimer J M 2012
{\it Ann. Rev. Nucl. Part. Sci.}  {\bf 62} 485.

\end{thebibliography}
\end{document}